\global\def\draftcontrol{0}

   \def\versionno{ gubser-mitra counterexample }

\catcode`\@=11

\expandafter\ifx\csname draftcontrol\endcsname\relax\global\def\draftcontrol{0}
\fi

{\count255=\time\divide\count255 by 60
\xdef\hourmin{\number\count255}
\multiply\count255 by-60\advance\count255 by\time
\xdef\hourmin{\hourmin:\ifnum\count255<10 0\fi\the\count255}}
\def\draftdate{\number\month/\number\day/\number\year\ \ \ \hourmin }

\newcommand\makepapertitle{\par
  \begingroup
    \renewcommand\thefootnote{\@fnsymbol\c@footnote}%
    \def\@makefnmark{\rlap{\@textsuperscript{\normalfont\@thefnmark}}}%
    \long\def\@makefntext##1{\parindent 1em\noindent
            \hb@xt@1.8em{%
                \hss\@textsuperscript{\normalfont\@thefnmark}}##1}%
     \newpage
     \global\@topnum\z@   
     \@makepapertitle
     \thispagestyle{empty}\@thanks
  \endgroup
  \setcounter{footnote}{0}%
  \global\let\thanks\relax
  \global\let\makepapertitle\relax
  \global\let\@makepapertitle\relax
  \global\let\@thanks\@empty
  \global\let\@author\@empty
  \global\let\@date\@empty
  \global\let\@title\@empty
  \global\let\title\relax
  \global\let\author\relax
  \global\let\date\relax
  \global\let\and\relax
  \def\version{\let\version\@version\@gobble}
}
\def\@makepapertitle{%
  \newpage
   \ifnum\draftcontrol=1 {}
   \version\versionno
   \vskip 3em%
   \else
   \hfill\hbox to 3cm {\parbox{4cm}{\@pubnum}\hss}%
   \vskip 3em%
   \fi
   \begin{center}%
   \let \footnote \thanks
     {\LARGE {\@title}}%
     \vskip 1.5em%
     {\normalsize
       \lineskip .5em%
       \begin{tabular}[t]{c}%
         \@author
       \end{tabular}\par}%
     \vskip 1.5em%
     {\@bstract}%
     \end{center}%
     \vskip 1.5em
     \@date%
   \par
}

\gdef\@pubnum{}
\def\pubnum#1{%
  \gdef\@pubnum{#1}}

\gdef\@bstract{}
\def\Abstract#1{%
  \gdef\@bstract{%
   \parbox{\textwidth-0pc}{%
   \centerline{\bf Abstract}\penalty1000%
\kern.2cm%
\noindent
\renewcommand\baselinestretch{1.0}%
{#1}}}
}

\def\ps@paper{\let\@mkboth\@gobbletwo%
     \ifnum\draftcontrol=1
    \def\@oddfoot{\hbox to \textwidth{\tiny \versionno \hfil\tiny\draftdate}%
    \hskip -\textwidth \hbox to \textwidth{\hfil\rm\thepage\hfil}}%
     \else\def\@oddfoot{\hbox to \textwidth{\hfil\rm\thepage\hfil}}
     \fi
     \let\@evenfoot\@oddfoot
}

\def\body{\clearpage
          \pagestyle{paper}
    }

\def\@version#1{\ifnum\draftcontrol=1
\typeout{}\typeout{#1}\typeout{}
\vskip3mm\centerline{\hbox{\fbox{\normalsize{\tt DRAFT -- #1 -- }
                   {\draftdate}}}}\vskip3mm
\fi}
\let\version\@version
\long\def\eqlabel#1{\ifnum\draftcontrol=1
                    \tag@false  
                    \tag*{(\theequation) \hbox to -0.2cm{\hspace{0cm}\small{#1}\hss}}
                    \refstepcounter{equation}
                    \edef\@currentlabel{\theequation}
                    \ltx@label{#1}          
                    \else
                    \label{#1}
                    \fi
                    }
\let\st@bibitem\@bibitem
\let\st@lbibitem\@lbibitem
\ifnum\draftcontrol=1
  \def\@bibitem#1{%
    \st@bibitem{#1}\a@@label{#1}\ignorespaces}
  \def\@lbibitem[#1]#2{%
    \st@lbibitem[#1]{#2}\a@@label{#2}\ignorespaces}
  \def\a@@label#1{%
    \gdef\a@lab{\smash{\normalfont\small#1}}
    \ifvmode
      \if@inlabel
        \global\setbox\@labels\hbox{%
          \llap{\a@lab\let\a@lab\relax
                \kern\@totalleftmargin\kern\marginparsep}%
          \box\@labels}%
      \fi
    \fi}
\fi

\documentclass[12pt,letterpaper]{article}

\usepackage{amsmath,amssymb,array,calc,epsfig,rotating,bm}
\usepackage[sort]{cite}
\usepackage{graphicx}
\usepackage{psfrag,verbatim}

\ifnum\draftcontrol=1
\fi

\renewcommand\baselinestretch{1.25}
\renewcommand\section{\@startsection {section}{1}{\z@}%
                                   {-3.5ex \@plus -1ex \@minus -.2ex}%
                                   {2.3ex \@plus.2ex}%
                                   {\normalfont\large\bfseries}}
\renewcommand\subsection{\@startsection{subsection}{2}{\z@}%
                                   {-3.25ex\@plus -1ex \@minus -.2ex}%
                                   {1.5ex \@plus .2ex}%
                                   {\normalfont\normalsize\bfseries}}
\renewcommand\subsubsection{\@startsection{subsubsection}{3}{\z@}%
                                   {-3.25ex\@plus -1ex \@minus -.2ex}%
                                   {1.5ex \@plus .2ex}%
                                   {\normalfont\normalsize\it}}
\renewcommand\paragraph{\@startsection{paragraph}{4}{\z@}%
                                   {-3.25ex\@plus -1ex \@minus -.2ex}%
                                   {1.5ex \@plus .2ex}%
                                   {\normalfont\normalsize\bf}}

\numberwithin{equation}{section}

\def\revise#1       {\raisebox{-0em}{\rule{3pt}{1em}}%
                     \marginpar{\raisebox{.5em}{\vrule width3pt\
                     \vrule width0pt height 0pt depth0.5em
                     \hbox to 0cm{\hspace{0cm}{%
                     \parbox[t]{4em}{\raggedright\footnotesize{#1}}}\hss}}}}

\def\cala         {{\cal A}}

\def\cale         {{\cal E}}
\def\calf         {{\cal F}}

\def\call         {{\cal L}}

\def\caln         {{\cal N}}
\def\calo         {{\cal O}}

\def\zet          {{\mathbb Z}}

\def\del          {\partial}

\def\Im           {{\rm Im\hskip0.1em}}

\def\sqr#1#2{{\vcenter{\vbox{\hrule height.#2pt
 \hbox{\vrule width.#2pt height#1pt \kern#1pt
 \vrule width.#2pt}\hrule height.#2pt}}}}

\newcommand{\qq}{\mathfrak{q}}
\newcommand{\ww}{\mathfrak{w}}
\newcommand{\hhh}{\mathbb{H}}

\newcommand{\beq}{\begin{equation}}
\newcommand{\eeq}{\end{equation}}
\newcommand{\beqa}{\begin{eqnarray}}
\newcommand{\eeqa}{\end{eqnarray}}
\newcommand{\beqar}{\begin{eqnarray*}}
\newcommand{\eeqar}{\end{eqnarray*}}

\renewcommand{\eqref}[1]{(\ref{#1})}

\newcommand{\ie}{{\it i.e.,}\ }

\def\r{\rho}

\def\dd{{\delta}}

\def\k{\kappa}
\def\e{\cale}
\catcode`\@=12

\begin{document}


\title{\bf Correlated stability conjecture revisited}
\pubnum
{UWO-TH-10/8
}

\date{October 2010}

\author{
Alex Buchel$ ^{1,2}$ and Chris Pagnutti$ ^{1}$\\[0.4cm]
\it $ ^1$Department of Applied Mathematics\\
\it University of Western Ontario\\
\it London, Ontario N6A 5B7, Canada\\
\it $ ^2$Perimeter Institute for Theoretical Physics\\
\it Waterloo, Ontario N2J 2W9, Canada\\
}

\Abstract{
Correlated stability conjecture (CSC) proposed by Gubser and Mitra
\cite{gm1,gm2} linked the thermodynamic and classical (in)stabilities 
of black branes. The classical instabilities, whenever occurring, were
conjectured to arise as Gregory-Laflamme (GL) instabilities of
translationally invariant horizons. In \cite{b} it was shown that the
thermodynamic instabilities, specifically the negative specific heat,
indeed result in the instabilities in the hydrodynamic spectrum of
holographically dual plasma excitations.  A counter-example of CSC was
presented in the context of black branes with scalar hair undergoing a
second-order phase transition \cite{ccsc}.  In this paper we discuss a
related counter-example of CSC conjecture, where a thermodynamically
stable translationally invariant horizon has a genuine tachyonic
instability. We study the spectrum of quasinormal excitations of a
black brane undergoing a continuous phase transition, and explicitly
identify the instability. We compute the critical exponents of the
critical momenta and the frequency of the unstable fluctuations and
identify the dynamical critical exponent of the model.
}

\makepapertitle

\body

\version\versionno
\tableofcontents

\section{Introduction}
Black holes with translationally invariant horizons (also referred to as black branes) 
are ubiquitous in string theory \cite{pol}. 
They are particularly important in the context of 
holographic gauge theory/string theory correspondence \cite{m9711}, where they 
describe the thermal equilibrium states of the dual gauge theory plasma. 
In case of $d=p+1$ space-time dimensional 
non-conformal gauge theory plasma, the dual black branes of the 
$(d+1)$-dimensional supergravity are rather complicated, and have multiple 
scalar hair, see for example \cite{bn2,bks}.
Being dual to a finite temperature gauge theory plasma, these black branes 
have nonvanishing Hawking temperature and necessarily are non-supersymmetric. 
Thus, their classical stability is not assured. The linear stability 
analysis of black branes is complicated by mixing the scalar hair and the 
metric fluctuations. A much simpler exercise is to compute the 
thermodynamic potentials of the black branes, \ie the free energy density, the 
energy density, etc., and analyze a thermodynamic stability  of the resulting 
potentials. It was a very welcome conjecture by Gubser and Mitra \cite{gm1,gm2}
---   {\it Correlated Stability Conjecture} --- that identified 
the thermodynamic stabilities or instabilities of translationally invariant 
horizons with the classical stabilities or instabilities correspondingly. 
Moreover, the classical instabilities, whenever occurring, were conjectured 
to appear as Gregory-Laflamme instabilities \cite{gl1,gl2}. 
Recall that GL instabilities of the translationally invariant horizons 
are linearized fluctuations $\delta \Phi\propto e^{-i \omega t +i\vec{k}\cdot\vec{x}}$
with a dispersion relation $(\ww,\qq)$ such that 
\begin{equation}
\Im(\ww)\bigg|_{\qq<\qq_c}\ >\ 0\,,
\eqlabel{disprel}
\end{equation}
where 
\begin{equation}
\ww\equiv \frac{\omega}{2\pi T}\,,\qquad \qq\equiv \frac{|\vec{k}|}{2\pi T}\,, 
\eqlabel{defwq}
\end{equation}
and $T$ is the temperature.
Note that we always assume that $\Im(\qq)=0$, \ie 
fluctuation amplitudes do not  grow exponentially along the translational directions 
of a horizon. 

A standard claim in classical thermodynamics\footnote{Assuming that the temperature 
is positive.} is that a system is thermodynamically stable if the Hessian $\hhh^\cale_{s,Q_A}$ of the energy 
density $\cale=\cale(s,Q_A)$ with respect to the entropy density $s$ and charges 
$Q_A\equiv\{Q_1,\cdots Q_n\}$, \ie
\begin{equation}
\hhh^\cale_{s,Q_A}\equiv \left(
\begin{array}{cc}
\frac{\del^2 \cale}{\del s^2} & \frac{\del^2\cale}{\del s\del Q_B}  \\
\frac{\del^2\cale}{\del Q_A\del s} & \frac{\del^2 \cale}{\del Q_A\del Q_B}  \end{array}
\right)\,,
\eqlabel{defhess}
\end{equation}
does not have negative eigenvalues. In the simplest case $n=0$, \ie no conserved charges, the thermodynamic stability 
implies that 
\begin{equation}
0\ <\ \frac{\del^2 \cale}{\del s^2}=\frac{T}{c_v}\,, 
\eqlabel{n0case}
\end{equation}
that is the specific heat $c_v$ is positive. In the context of gauge theory/string theory 
correspondence black holes with translationary invariant horizons in asymptotically anti-de-Sitter space-time 
are dual (equivalent) to equilibrium 
thermal states of certain strongly coupled systems. Thus, the above thermodynamic stability criteria should be 
directly applicable to black branes as well. CSC asserts that it is only when the Hessian \eqref{defhess}
for a given black brane geometry is positive, the spectrum of on-shell excitations in this background geometry 
is free from tachyons.

CSC is trivially true in a holographic context, whenever the 
specific heat of the corresponding black brane is negative \cite{b}. Indeed,
in the absence of conserved charges, the speed of hydrodynamic (sound channel)
modes in the dual plasma is  
\begin{equation}
c_s^2=\frac{s}{c_v}\,,
\eqlabel{vs2}
\end{equation} 
where $s$, $\cale$ are the entropy and the energy densities, and 
$c_v$ 
\begin{equation}
c_v =\left(\frac{\del\cale}{\del T}\right)_v\,,
\eqlabel{cvdef}
\end{equation}
is the specific heat. Thus whenever $c_v<0$, the sound modes in 
plasma, and correspondingly the dual quasinormal modes in the gravitational 
background, are unstable\footnote{Such instability occurs in $\caln=2^*$ holographic 
plasma \cite{bn2,bp4}.}. 

The situation is more complicated in holographic examples in the presence 
of conserved charges. Consider a strongly coupled $\caln=4$ $SU(N)$ supersymmetric 
Yang-Mills theory in the planar limit at finite temperature $T$ and a chemical 
potential $\mu$ for a single global $U(1)\subset SO(6)$ R-symmetry. 
The dual gravitational geometry is that of the Reissner-Nordstr\"om asymptotic 
$AdS_5$ black brane \cite{stu1,stu2,stu3}. 
For any  temperature larger than the critical one, $T_c=\mu \sqrt{2}/\pi$, 
there are two phases of the plasma. Although one of the phases has a negative 
specific heat, the sound modes in plasma are always stable \cite{bn4}:
\begin{equation}
c_v=4\pi^2N^2T^3\frac{(1+\k)(3-\k)}{(2+\k)^2(2-\k)}
\,,\qquad \ww=\pm \frac{1}{\sqrt{3}}\ \qq - \frac{i}{3}\ \frac{\k+2}{2\k+2}\ 
\qq^2+\calo(\qq^3) \,,
\eqlabel{sound}
\end{equation} 
where 
\begin{equation}
 \frac{2\pi T}{\mu}=\sqrt{\k}+\frac{2}{\sqrt{\k}} \,,
\eqlabel{defk}
\end{equation}
so that given $T>T_c$ there are two possible values of $\k$ distinguishing 
two different phases: one with $\k<2$ (a thermodynamically stable phase) 
and one with $\k>2$
(a thermodynamically unstable phase). 
There is no contradiction here with the argument presented in \cite{b} ---
the speed of sound in the presence of conserved charges takes a more complicated form 
than in \eqref{vs2}. In case of a charged  plasma we have 
\begin{equation}
c_s^2=\biggl((\e+P)\ \frac{\del(P,\r)}{\del(T,\mu)}
+\r\ \frac{\del(\e,P)}{\del(T,\mu)}\biggr)\biggl((\e+P)\ \frac{\del(\e,\r)}
{\del(T,\mu)}\biggr)^{-1}\,,
\eqlabel{cs2mu}
\end{equation}
where $P$ is the pressure. 
Remarkably, there is no contradiction with CSC conjecture as well! 
Indeed, it was shown in \cite{bn4} that a two-point retarded correlation function 
of the charge density fluctuations has a pole (and thus the SYM plasma 
has a physical excitation with this dispersion relation) at 
\begin{equation}
\ww=\cala\ i (\k-2)\ \qq^2+\calo(\qq^4)\,,\qquad \cala=0.33339(2) \,,
\eqlabel{pole}
\end{equation}
so that a phase with $\k>2$ is classically unstable, at least 
with respect to fluctuations with sufficiently 
small $\qq$. We emphasize that even though the gravitational instability is 
in the quasinormal sound channel mode, it is not a hydrodynamic sound, 
which is characterized by the dispersion relation $\ww\propto \qq$, see \eqref{sound}.

So far, the examples we discussed provide  support for CSC conjecture. It is 
believed however, that CSC conjecture is false 
\cite{ccsc}\footnote{See also \cite{m}.}. The idea advocated in
\cite{ccsc} is to consider gravitational backgrounds with translationally invariant 
horizon that are dual to  gauge theory plasmas undergoing a continuous phase transition.
In the vicinity of the phase transition the condensate does not noticeably 
modify the thermodynamics, and thus should not affect the thermodynamic stability
of the system. On the other hand, the phase of the system with the higher 
free energy is expected to be classically unstable. The condensation of the tachyon 
should bring the system to the equilibrium phase with the lowest free energy. 
The tachyon of the unstable plasma phase should appear 
in a dual gravitational description as a  tachyonic quasinormal mode, \ie 
the GL instability \eqref{disprel}.

In this paper we discuss an explicit example of the physical scenario
suggested in \cite{ccsc} from the perspective of the quasinormal spectrum. 
We identify the GL tachyon and compute the critical exponents for the critical momentum
and the frequency of the corresponding fluctuations. Our model is the 'exotic 
hairy black hole' introduced in \cite{bp2}. The dispersion relation for the 
sound waves in  this model was studied in \cite{bp3}. Much like in case 
of the R-charged $\caln=4$ SYM plasma, there are no instabilities 
in hydrodynamic modes. We review the thermodynamics and the hydrodynamics of the model 
in section 2. In section 3 we identify non-hydrodynamic quasinormal excitations 
in the sound channel of the gravitational background with the dispersion 
features identical to those of the GL instabilities \eqref{disprel}. 
The GL tachyons exist in the symmetry broken phase only, and disappear from the spectrum
at the second-order phase transition. We compute critical exponents associated with 
vanishing of $\qq_c$ and $\ww_c\equiv \ww(\qq=0)$ at the transition.
Finally, we comment on the dynamical universality class of the model.   
    
\section{Thermodynamics and hydrodynamics of exotic black branes}
The model considered here is not a string theory derived example of gauge/gravity correspondence ---
rather, it should be viewed as a phenomenological model of holography. Of 
course, whether or not 
the model can be embedded into the full string theory is irrelevant in so far as one is 
interested in CSC. The holography here is simply a useful tool to think physically 
about mathematical problem of (in)stabilities of translationally invariant horizons.

Following \cite{gubser}, consider a relativistic conformal field theory in 
2+1 dimensions, deformed by a relevant operator $\calo_r$:
\begin{equation}
H_{CFT}\to \tilde{H}=H_{CFT}+\lambda_r \calo_r\,.
\eqlabel{ham3d}
\end{equation}
Such a deformation softly breaks the scale invariance and induces the
renormalization group flow.  We further assume that the deformed
theory $\tilde{H}$ has an irrelevant operator $\calo_i$ that mixes
along the RG flow with $\calo_r$. The explicit holographic model
realizing this scenario was discussed in \cite{bp2}:
\begin{equation}
\begin{split}
S_4=&S_{CFT}+S_{r}+S_i=\frac{1}{2\kappa^2}\int dx^4\sqrt{-\gamma}\left[\call_{CFT}+\call_{r}+\call_i\right]\,,
\end{split}
\eqlabel{s4}
\end{equation}
\begin{equation}
\call_{CFT}=R+6\,,\qquad \call_r=-\frac 12 \left(\nabla\phi\right)^2+\phi^2\,,\qquad 
\call_i=-\frac 12 \left(\nabla\chi\right)^2-2\chi^2-g \phi^2 \chi^2 \,,
\eqlabel{lc}
\end{equation}
where we split the action into (a holographic dual to)  a CFT part $S_{CFT}$; its deformation by a relevant
operator $\calo_r$; and a sector $S_i$ involving an irrelevant operator $\calo_i$ 
along with its mixing with $\calo_r$ under the RG dynamics.
The four dimensional gravitational constant $\kappa$ is related to the central charge $c$ of the 
UV fixed point as 
\begin{equation}
c=\frac{192}{ \kappa^2}\,.
\eqlabel{cg}
\end{equation}
In our case the scaling dimension of $\calo_r$ is 2  
and the scaling dimension of $\calo_i$ is 4.  In order to have asymptotically $AdS_4$ solutions, 
we assume that only the normalizable mode of $\calo_i$ is nonzero near the boundary.
Finally, we assume that $g<0$ in order to holographically induce the critical behavior 
\cite{gubser}.

\begin{figure}[t]
\begin{center}
\psfrag{Fred}{{$\frac{\Omega}{(\pi T)^3}$}}
\psfrag{TcT}{{$\frac{T_c}{T}$}}
\psfrag{O}{{$\langle\calo_i\rangle^2$}}
\includegraphics[width=3in]{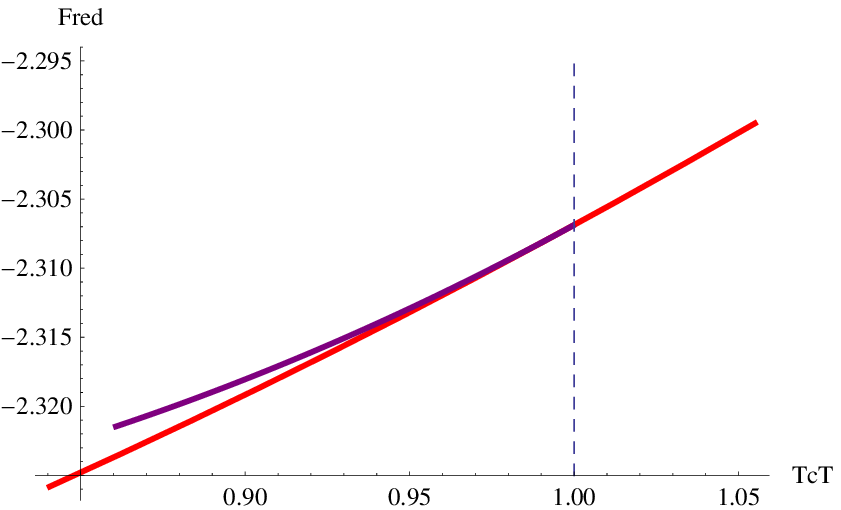}
\includegraphics[width=3in]{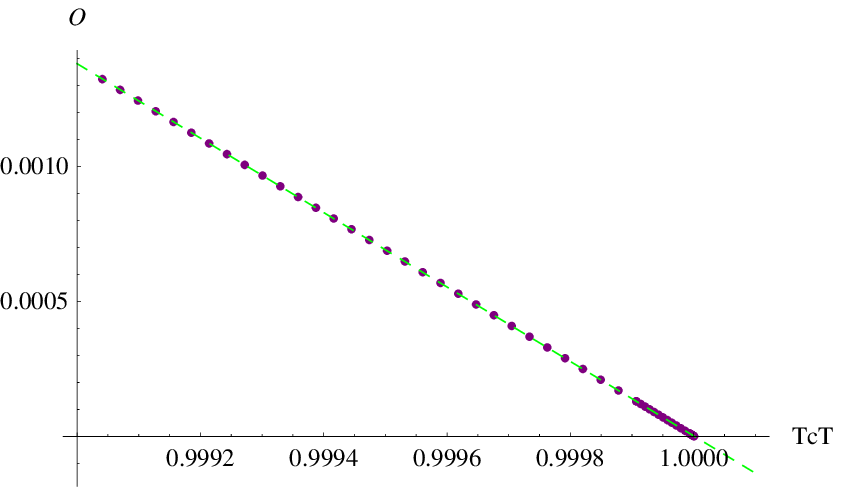}
\end{center}
  \caption{(Colour online)
The free energy densities  $\Omega_0$ of the symmetric phase (red curve, left plot) and $\Omega_d$ of 
the symmetry broken 
phase (purple curve, left plot) as a function of the reduced temperature $\frac{T}{T_c}$ in the gauge theory plasma dual 
to the holographic RG 
flow in \cite{bp2}. The right plot represents the square of $\langle\calo_i\rangle$  
(which we use as an order parameter for the transition) as a function 
of the reduced temperature. The dashed 
green line is a linear fit to  $\langle\calo_i\rangle^2$.
 } \label{figure1}
\end{figure}

\begin{figure}[t]
\begin{center}
\psfrag{cs}{{$2 c_s^2$}}
\psfrag{tt}{{$\frac{T_c}{T}$}}
\psfrag{zetaeta}{{$\frac{\zeta}{\eta}$}}
\includegraphics[width=3in]{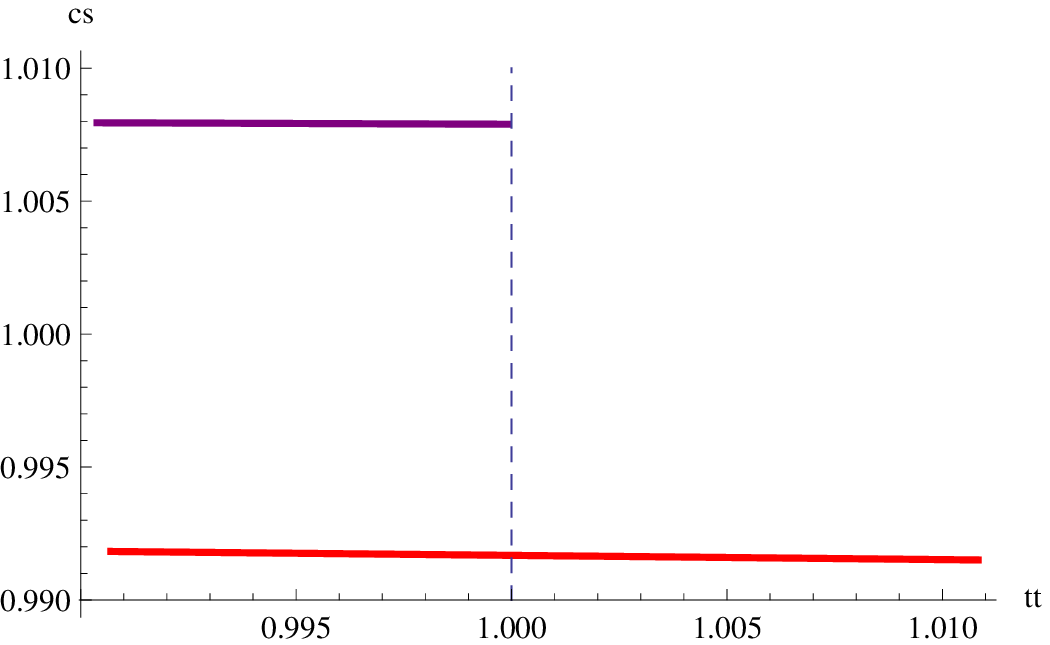}
\includegraphics[width=3in]{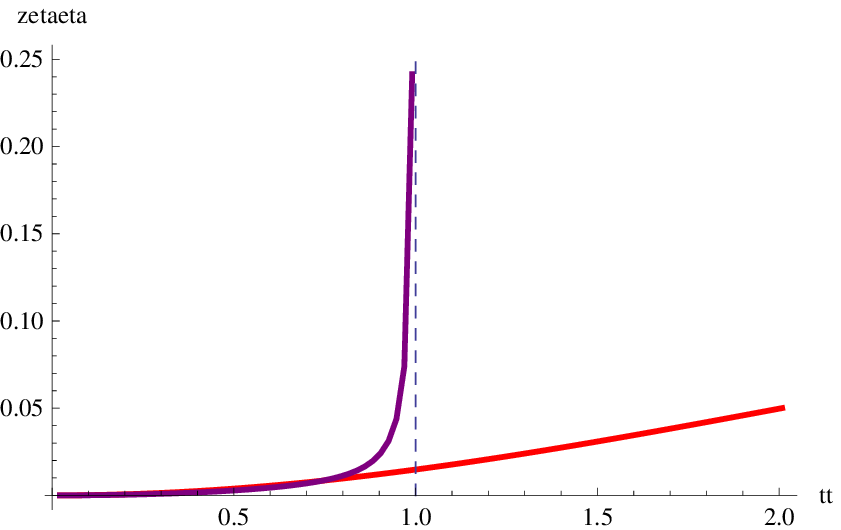}
\end{center}
  \caption{(Colour online)
The speed of sound $c_s$ and the ratio of bulk-to-shear viscosities  
in the gauge theory plasma dual to the holographic RG 
flow in \cite{bp2} as a function of the reduced temperature $\frac{T_c}{T}$.
 }\label{figure2}
\end{figure}

Effective action \eqref{s4} has a $\zet_2\times \zet_2$ discrete symmetry that acts as a parity transformation on 
the scalar fields $\phi$ and $\chi$. The discrete symmetry $\phi\to -\phi$ is softly broken by a relevant deformation 
of the $AdS_4$ CFT; while the $\chi\to -\chi$ symmetry is broken spontaneously. 
Spontaneous breaking of the latter symmetry is caused by the development of the condensate 
for the irrelevant operator $\calo_i$. The unusual part of this phase transition is
that the symmetry broken phase occurs at high temperatures (rather
than at low temperatures) and that the broken phase has a higher free energy
density than the unbroken phase with $\langle\calo_i\rangle=0$. The free energy 
densities of the $\zet_2$-symmetric phase ($\langle \calo_i \rangle=0$) and the symmetry 
broken phase  ($\langle \calo_i \rangle\ne 0$) are presented in Figure \ref{figure1}.

The hydrodynamic sound channel quasinormal modes of both phases were studied in details in
\cite{bp3}. There is a jump discontinuity at the transition in the speed of sound; with the 
speed of sound in the symmetry broken phase being $\sim 1\%$ higher than in the symmetric 
phase. Both the shear and the bulk viscosities in the two phases are positive. 
Figure \ref{figure2} collects the results for the speed of 
sound and the ratio of bulk-to-shear viscosities in the exotic plasma.  In the symmetry broken 
phase the bulk viscosity diverges in the vicinity of the phase transition 
\cite{bp3}\footnote{The shear viscosity is continuous and finite at the transition: 
$\eta/s=1/4\pi$ \cite{sh1,sh2,sh3}.}
\begin{equation}
\frac{\zeta}{\eta}\bigg|_{broken}\ \propto\ |\langle\calo_i\rangle|^{-2}\ \propto\ (T-T_c)^{-1} \,.
\eqlabel{bulk}
\end{equation}

Note that since $c_s^2>0$, the specific heat (see \eqref{vs2}) of both the symmetric and the 
symmetry broken phases is positive --- the two phases are thermodynamically 
stable\footnote{We also verified that the susceptibility $\frac{\del^2 \cale}{\del \Lambda^2}\bigg|_{s=constant}$ 
is positive --- $\Lambda$ is the scale introduced by the relevant operator $\calo_r$, 
see \cite{bp2} for more details.}. 
The dispersion relation for the sound mode takes the form
\begin{equation}
\ww=\pm c_s\ \qq-\frac{i}{4}\ \left(1+\frac{\zeta}{\eta}\right)\ \qq^2+\calo(\qq^3)\,,
\eqlabel{sound3d}
\end{equation}
implying that the sound modes are stable as well. 

Since the free energy of the symmetry broken phase is bigger, this phase is not 
thermodynamically preferable and one expects a classical instability driving the system
to a true (symmetric) ground state. Such instability is expected to be of GL type,
see \eqref{disprel}, and thus is not expected to show up in the hydrodynamic limit.
In the next section we explicitly identify the Gregory-Laflamme instability,
and thus prove that CSC  is false.

\section{GL tachyons of exotic black branes}

\begin{figure}[t]
\begin{center}
\psfrag{qq}{{$\qq^2$}}
\psfrag{tt}{{$\frac{T_c}{T}$}}
\includegraphics[width=3in]{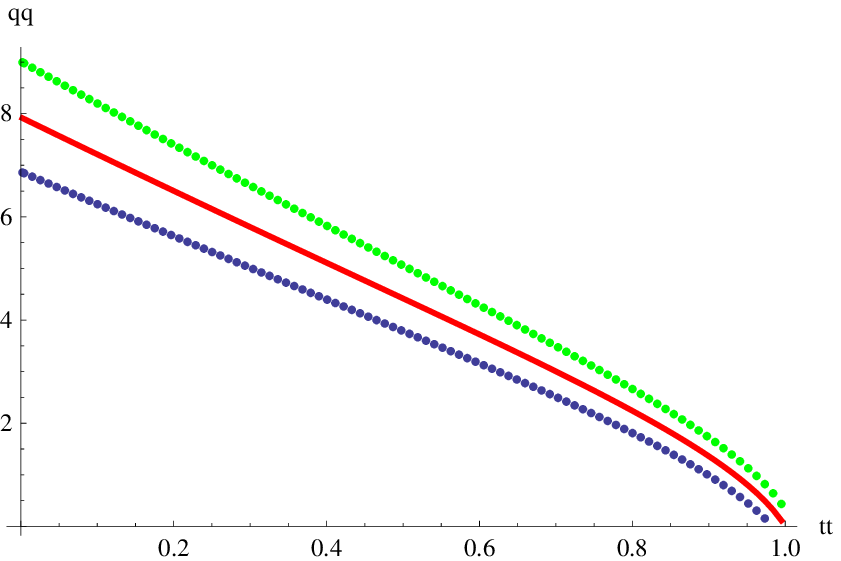}
\includegraphics[width=3in]{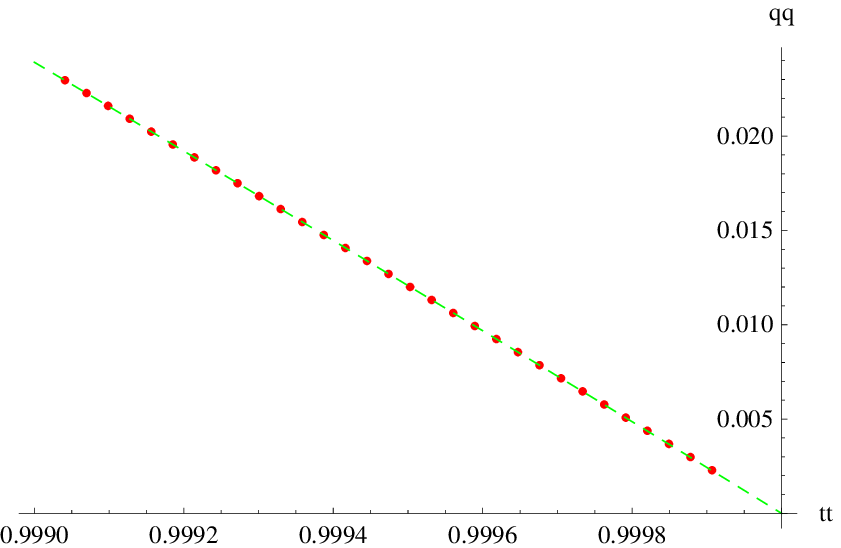}
\end{center}
  \caption{(Colour online) Dispersion relation of quasinormal modes in the symmetry broken (classically 
unstable)
phase of the exotic black branes. The red line (left plot) represents the modes at the 
threshold of GL instability: $(\ww,\qq)=(0,\qq_c)$. The green/blue dots (left plot) 
are the stable/unstable quasinormal 
modes with $\ww=- 0.1 i$ and $\ww= 0.1 i$ correspondingly. The right plot presents the modes 
at the threshold of instability in the vicinity of the critical point. The dashed green line
(right plot) is the best quadratic in $\left(1-\frac {T_c}{T}\right)$ fit to data. 
 } \label{figure3}
\end{figure}

\begin{figure}[t]
\begin{center}
\psfrag{iw}{{$i\ww_c$}}
\psfrag{tt}{{$\frac{T_c}{T}$}}
\includegraphics[width=3in]{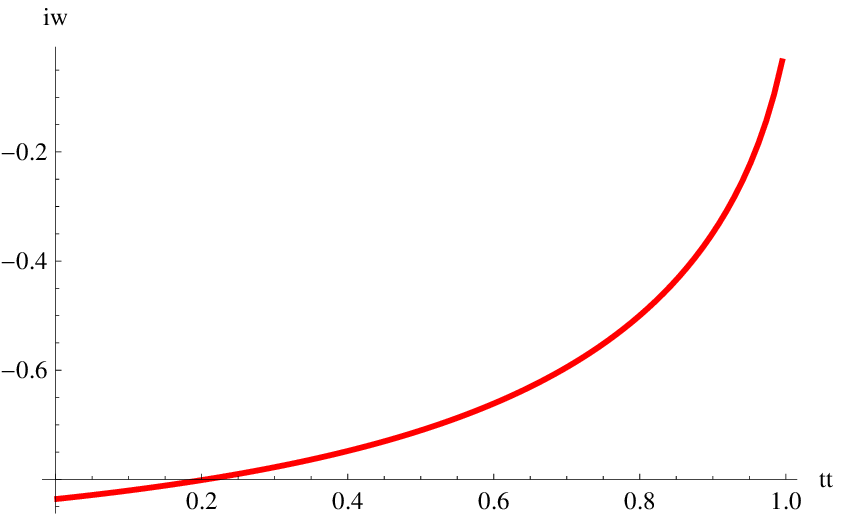}
\includegraphics[width=3in]{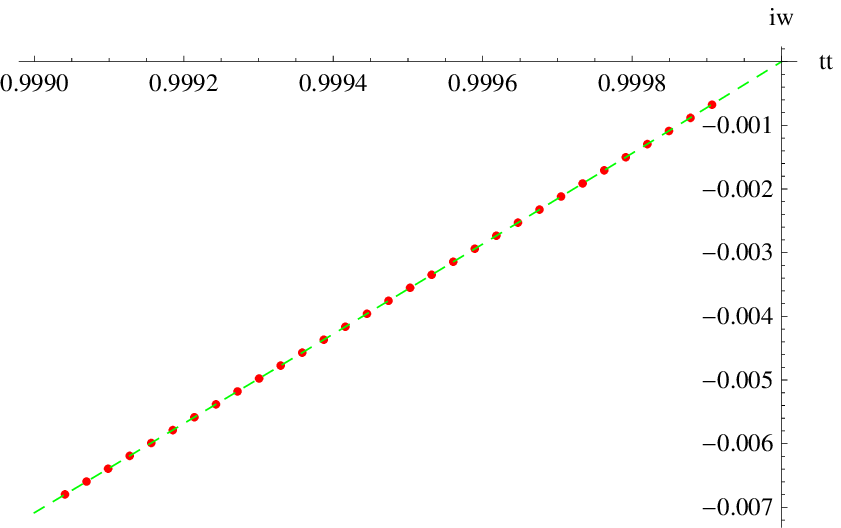}
\end{center}
  \caption{(Colour online) Frequency of the homogeneous and isotropic unstable quasinormal modes
$(\ww,\qq)=(\ww_c,0)$ 
in the symmetry broken (classically unstable) phase  of the exotic black branes.  The dashed green line
(right plot) is the best quadratic in $\left(1-\frac {T_c}{T}\right)$ fit to data 
in the vicinity of the critical point. 
 } \label{figure4}
\end{figure}

Physical excitations in a gauge theory plasma are dual to quasinormal modes of a 
corresponding black brane solution \cite{kovs}. For a translationally invariant horizon, 
a wave-function of a generic background field fluctuation $\dd\Phi$ takes form 
\begin{equation}
\dd\Phi=\calf(r) e^{-i\omega t+i\vec{k}\cdot \vec{x}}\,,
\eqlabel{wave}
\end{equation}
where $\calf$ is a radial wave function. Fluctuations of different fields will typically 
mix with each other. It is convenient to decompose the fluctuations in irreducible representations 
with respect to rotations about the $\vec{k}$-axis. In the case of the exotic black branes, 
the GL instabilities are expected to arise as fluctuations in the symmetry-breaking (scalar) condensate. 
Thus, they appear in the {\it scalar} quasinormal mode channel. It is exactly the same set of 
fluctuations which in the hydrodynamic limit ($\ww\to 0$, $\qq\to 0$ and $\frac \ww\qq\to {\rm constant}$)
describe the propagation of the sound waves in plasma.  
The technique for computing the scalar channel quasinormal modes 
in non-conformal holographic models was developed in \cite{bbs}. 
 It is straightforward to generalize the method of \cite{bbs} to the computation of the scalar channel 
quasinormal modes of 
the gauge theory plasma dual to the holographic RG flow \cite{bp2}. 
The details of the latter analysis will appear elsewhere \cite{pph} 
and here we report only the results, focusing on the GL 
mode\footnote{Sound waves were discussed in \cite{bp3} and reviewed in the previous section.}. 

The red solid line (left plot) in Figure \ref{figure3} presents the dispersion relation for the
quasinormal modes in the symmetry broken (classically unstable) phase at the threshold of instability:
\begin{equation}
(\ww,\qq)\bigg|_{threshold}\ =\ (0,\qq_c)\,,
\eqlabel{thr}
\end{equation} 
as a function of the reduced temperature $\frac{T_c}{T}$.
The modes with $\qq^2> \qq_c^2$ attenuate, \ie they have $\Im(\ww)<0$, while those 
with $\qq^2< \qq_c^2$ represent a genuine GL instability --- they have $\Im(\ww)>0$.
The green dots on the left plot have $\ww=-0.1 i $ and the blue dots have 
$\ww=0.1 i$. The right plot on Figure \ref{figure3} represents the dispersion 
relation of the quasinormal modes in the 
symmetry broken phase at the threshold of instability in the vicinity of the 
critical point. The green dashed line is the best quadratic in $t\equiv (1-\frac{T_c}{T})$
fit to the data:
\begin{equation}
\qq_c^2\bigg|_{fit}=3.(4)\times 10^{-5}+24.(3)\ t-369.(1)\ t^2+\calo(t^3)\,.
\eqlabel{greenq}
\end{equation} 
The data suggests that 
\begin{equation}
\qq_c^2\propto (T-T_c)\,,
\eqlabel{qc}
\end{equation}
in the vicinity of the critical point.

The left plot in Figure \ref{figure4} represents the unstable homogeneous and
isotropic quasinormal modes, \ie GL instabilities with $\qq=0$. The right plot 
shows the frequency dependence of these modes in the vicinity of the critical point.
The green dashed line is the best quadratic  in $t$ fit to the data:
\begin{equation}
i\ww_c\bigg|_{fit}=-9.(8)\times 10^{-6}-7.(2)\ t+102.(9)\ t^2+\calo(t^3)\,.
\eqlabel{greenw}
\end{equation} 
The data suggests that 
\begin{equation}
\ww_c\propto i (T-T_c)\,,
\eqlabel{wc}
\end{equation}
in the vicinity of the critical point.

Notice that since  $\ww_c\propto i \qq_c^2$ in the vicinity of the critical point,
it is natural to identify the dynamical critical exponent of the model with 
$z=2$. The symmetries of our model identify it as 'model A' according to classification 
of Hohenberg and Halperin (HH) \cite{hh}, assuming a natural extension of the classification 
to classically unstable phases. Mean-field models (with vanishing anomalous static critical 
exponent)
in the HH universality class A are predicted to
have the dynamical critical exponent $z_{model-A}=2$, as the one we 
obtained\footnote{The second-order phase transition in exotic black branes is of the 
mean-field type \cite{pph}.}.

In this section we explicitly identified the Gregory-Laflamme instabilities 
in the symmetry broken phase of the exotic 
black branes. Since this phase is thermodynamically stable, our model 
presents a counter-example to the  correlated stability conjecture.

\section*{Acknowledgments}
We would like to thank Martin Kruczenski for valuable discussions.
Research at Perimeter Institute is
supported by the Government of Canada through Industry Canada and by
the Province of Ontario through the Ministry of Research \&
Innovation. AB gratefully acknowledges further support by an NSERC
Discovery grant and support through the Early Researcher Award
program by the Province of Ontario.

\end{document}